\documentclass[12pt]{article}     
\usepackage[margin=1in,footskip=0.25in]{geometry}                                                                                                                                                
\usepackage{graphicx}   
\usepackage{amssymb}   
\usepackage{float}

\usepackage{color}  \color{blue}  \color{black}

\title{Third rank permeability in chiral solids} 
\author{Roderic S. Lakes\\
Department of Materials Science, Department of Engineering Physics\\ 
Department of Mechanical Engineering\\
University of Wisconsin\\
1500 Engineering Drive, Madison, WI 53706-1687}

\begin{document}

\maketitle
\color{blue} Preprint, Transport Phenomena 1 (1) (2026) \color{black}

\begin{abstract}
	  Effects of a third rank permeability term in chiral solids are studied. Fluid flow through such materials acquires vorticity upon emergence from the material. Materials of interest include chiral surface lattices such as the gyroid, chiral rib lattices, and granular materials comprised of sugar crystals, quartz sand, wheat or beans. A characteristic length scale is associated with the chirality. The length scale can be obtained by several methods. Contacts with nonlocal permeability, elasticity and piezoelectricity are explored. 
\end{abstract}


\section{Introduction}
Many properties of materials are describable by tensors  \cite{Nye} \cite{Lovett}. As for tensors of even rank, the heat capacity is represented by a scalar, a tensor of rank zero.  The thermal expansion and electrical and thermal conductivity are represented by tensors of rank two.  For thermal conductivity $k_{ij}$, with  $h_{i}$ as the thermal current and $E_{j} = T ,_{j}$ as the field of gradient of temperature $T$. 
\begin{equation} \label{eq:thermalcond2nd}
h_{i} = k_{ij} E_{j} 
\end{equation} 
in which the Einstein summation convention over repeated indices is used and the comma denotes differentiation with respect to spatial coordinates. Even rank tensors that govern dielectric permittivity, elasticity and magnetic permeability are symmetric based on   the assumption of a conserved energy density \cite{Nye}. Tensors for conductivity are taken to be symmetric \cite{Nye} via the Onsager relations \cite{Onsager1, Onsager2}. 
\par
The classical elastic modulus is represented by a tensor of rank four \cite{Sokolnikoff}. The tensor representation allows anisotropy but does not require it. For a classically elastic solid, the stress ${\sigma}_{ij}$  is related to the strain ${\epsilon}_{kl}$ via the elastic modulus tensor $C_{ijkl}$ 

\begin{equation} \label{eq:ElastClassic}
{\sigma}_{ij} = C_{ijkl}  {\epsilon}_{kl}
\end{equation}
in which the Einstein summation convention over repeated indices is used. Chirality has no effect in classical elasticity because the elastic modulus tensor is fourth rank. An inversion of coordinates used to invoke chirality has no effect on tensors of even rank so classical elasticity does not admit any effect of chirality. Also the modulus tensor is symmetric, $C_{ijkl} = C_{klij}$ in elastic solids for which there exists a conserved strain energy density. 
\par
There is a difference in the freedom allowed by even and odd rank tensors. Effects of chirality  (the material has left and right handed forms) in a material are manifest only in true tensor properties of odd rank \cite{Nye}. Classical piezoelectricity and pyroelectricity occur only in chiral materials. Piezoelectricity is quantified by a tensor of third rank; pyroelectricity by a tensor of rank 1. Piezoelectricity provides sensitivity of the electric displacement vector $\mathcal{D}_{i}$ to stress ${\sigma}_{ij}$ by the classical third rank direct effect piezoelectric coupling $d_{ijk}^{d}$ via $\mathcal{D}_{i}  = d_{ijk}^{d}   {\sigma}_{jk}$. Chirality is essential for this tensor to have nonzero elements and for classical piezoelectricity to occur. We consider only true tensors, not pseudo-tensors.  
\par
More freedom is allowed if higher rank tensor terms are added to terms that are classically known. 
For example, some piezoelectric materials are sensitive to the gradient of stress \cite{Bursian1968} as well as to stress ${\sigma}_{ij}$ as has long been known experimentally \cite{Bursian1974}  \cite{Trunov1975}. 

\begin{equation} \label{eq:piezograd}
\mathcal{D}_{i}  = d_{ijk}^{d}   {\sigma}_{jk} + d_{ijkl} \frac{{\partial} {\sigma}_{jk}}{{\partial} {x_{l}}}
\end{equation} 

Such effects involve a fourth rank coupling tensor term $d_{ijkl}$ as in Eq. \ref{eq:piezograd} in addition to the effect of the usual tensor of third rank. For fourth rank effects, the material need not be chiral;  effects can be large in ferroelectric materials and in materials near a phase transition. Recently such effects are called flexo -electric.
\par
  These effects can be subsumed as part of an expansion of a nonlocal  \cite{Bursian1968} functional in which the effect at a point is due to the cause not just at the point but in a region around that point.  This functional can be expanded in a series of higher terms as in in Eq. \ref{eq:nonlocal}. Nonlocal theories of elasticity \cite{EringenNonl72}, piezoelectricity \cite{Bursian1968} \cite{Bursian1974}  and permeability \cite{WaismanNonl17} are known. 
\begin{equation} \label{eq:nonlocal}
\mathcal{\mathbf{D}} = \int{{\mathbf{d}}(\mathbf{r - r'}) {\mathbf{\sigma}}(\mathbf{r'}) \mathbf{dr'}} = d_{ijk}^{d}   {\sigma}_{jk} + d_{ijkl} \frac{{\partial} {\sigma}_{jk}}{{\partial} {x_{l}}} + ...
\end{equation} 
The higher order terms involve true tensors of higher rank than that of the usual piezoelectricity tensor. 
 
\par
Gradient effects can be incorporated in elasticity by adding a term of fifth rank $C_{ijklm} \frac{{\partial} {\sigma}_{kl}}{{\partial} {x_{m}}}$  to the fourth rank modulus in Eq. \ref{eq:ElastClassic} or by incorporating an independent rotation variable for points as in Cosserat \cite{Cosserat09}  (micropolar \cite{Eringen68}) elasticity. In that case, the theory reduces to gradient sensitive elasticity if the Cosserat coupling coefficient is large enough that the local rotation of points is equal to the rotation associated with displacement gradients. Cosserat freedom has been demonstrated experimentally in many materials with a microstructure size that is nontrivial compared with the scale of the experiment e. g. bone \cite{ParkLakesBone86}, diamond crystals \cite{NagawaDiamond80},   foams \cite{RuLakesPhilM16} \cite{RuegLak16} and lattices of ribs \cite{RuegLakLatt18}. Chirality, though it has no effect in classical elasticity, gives rise to stretch - twist or squeeze - twist coupling as demonstrated theoretically \cite{CossChiral82} and experimentally \cite{HaLakChiral16}, \cite{ReasaLakChiral20}, \cite{HaLakChiral20}.
\par
Chirality is shown here via a third rank term to have an effect in fluid flow permeability through a porous solid. Flow in chiral media will give rise to circulating flow and vorticity via a third rank tensor.

\section{Analysis}
\begin{figure}[htb!]
\centering
\includegraphics[width=0.25\textwidth]{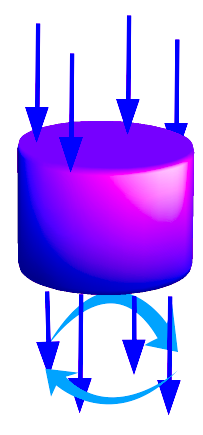} 
\caption{Fluid flow through a chiral permeable cylinder. Pressure $P$ is applied at the top surface. Arrows indicate flow direction. Flow exiting the bottom of the cylinder acquires vorticity as indicated by the curved arrows. }
\label{chiralflow}
\end{figure}

Conductive flow of heat through a solid and flow of fluid through a porous solid are governed by similar equations but there are differences in consequences, so both will be discussed. Heat flow is classically represented by a tensor of rank two as given in Eq. \ref{eq:thermalcond2nd}. 
If a third rank term $k_{ijk}$ is added we have the heat flow $h_{i}$

\begin{equation} \label{eq:thermalcond3rd}
h_{i} = k_{ij} E_{j} + k_{ijk} E_{j,k}
\end{equation} 
in which $E_{j}$ is the gradient  $T_{,j}$ of temperature $T$ and the comma represents differentiation with respect to spatial coordinates.  The heat flow due to a longitudinal temperature gradient in a bar via the third rank term follows a spiral path with a longitudinal and a circumferential component. The flow direction is ordinarily not observable but can be made observable by preparing a specimen with a notch so that circumferential flow, if present, results in a build up of heat near the notch. This manifests itself as a temperature difference and has been observed.

Consider flow of fluid through a block of porous solid \cite{GibAshby}. The flow velocity $v_{j}$ is related following Darcy's law to the gradient $P_{,j}  = P/H$ of pressure $P$ by the permeability $\kappa_{ij}$, with $H$ as the thickness of the porous block; $R$ is its radius. 

\begin{equation} \label{eq:classicpermeable}
v_{j}   = \kappa_{ij} P_{,j}
\end{equation}
The permeability may be written $\kappa_{ij} = K_{ij}/ \mu$ with $\mu$ as the fluid viscosity and $K_{ij}$ as absolute permeability proportional to the square of the pore size.

Suppose the flow velocity $v$ is given in terms of field of pressure $P$ via a superposition of contributions from a second rank tensor and a third rank tensor term $\kappa_{ijk}$. 
\begin{equation} \label{eq:permeable3rd}
v_{i} = \kappa_{ij} P_{,j} + \kappa_{ijk} P_{,jk}
\end{equation} 

It is expedient to consider an inverse form 
\begin{equation} \label{eq:permeable3rdInv}
 P_{,i} = \rho_{ij} v_{j} + \rho_{ijk} v_{j,k}
\end{equation} 
in which $\rho_{ij}$ is the flow resistivity and  $\rho_{ijk}$ is the third rank resistivity term. The resistivity terms have different dimensions due to the derivative. 

The material can be anisotropic but interpretation is facilitated if isotropy is assumed. Then $\rho_{ij} =  \rho_{L} \delta_{ij}$ with $\rho_{L}$ as the longitudinal resistivity and $\delta_{ij}$ as the Kronecker delta.  
There is one isotropic third rank tensor, the permutation symbol  (Levi Civita symbol)   $e_{ijk}$.  It is antisymmetric, $e_{ijk} = -e_{ikj}$. 
\par
So 
\begin{equation} \label{eq:permeable3rdInvIso}
P_{,i}  = \rho_{L} \delta_{ij} v_{j} + \rho_{T} e_{ijk} v_{j,k}
\end{equation} 
If one defines $\rho_{L}$ and $\rho_{T}$ to have the same dimensions then a factor of the radius $R$ may be introduced in the second term of Eq. \ref{eq:permeable3rdInvIso}. The block of material is assumed to be a cylinder of radius $R$ sealed with respect to radial outflow on the curved surfaces.
\par
Here, $\rho_{T}$ is a material constant for transverse flow and the second term contains curl $v$. Recall that curl $v_{i} = e_{ijk}v_{j,k}$.  The curl of a gradient equals zero.  However the derivatives of the current are not subsumed in a gradient when a third rank term is present. The derivative $v_{j,k}$ need not equal $v_{k,j}$. 
\par
Define a characteristic length $\ell_{p}$ with dimensions of length so that 
$\rho_{T} = \frac{\rho_{L}}{\ell_{p}}$ in which $\rho_{L}$ and $\rho_{T}$ are the longitudinal and transverse resistivity coefficients respectively. If one defines $\rho_{L}$ and $\rho_{T}$ to have the same dimensions, then the characteristic length is expressed $\rho_{T} = \frac{R}{\ell_{p}}\rho_{L}$. 
\par
Apply Green's theorem, $\int curl \mathbf{v} dA =  \oint \mathbf{v} \cdot \mathbf{dl}$, to the curl term in Eq. \ref{eq:permeable3rdInvIso}, so for a circular contour of radius $r$ $P_{L}/H = \rho_{L}v_{L} + 2 \rho_{T} v_{T}(r)$.
Using the definition of characteristic length, the nonclassical transverse circulating flow at the surface $r = R$ is
\begin{equation} \label{eq:permeable3rdTr}
v_{T} = (P_{L}/H  -  \rho_{L}v_{L} ) \frac{\ell_{p}} {R} \frac{1} {2\rho_{L}} 
\end{equation}
so the nonclassical flow is proportional to a characteristic length.  Thicker cylinders of larger radius are expected to exhibit a smaller transverse flow than thinner ones. Characteristic length effects are also known in the rigidity predicted and observed in nonclassical (Cosserat) elasticity and piezoelectricity as referenced above.   The example of isotropy is not essential. Anisotropic materials will have more physical constants and will have more freedom.  
\par
The second term in Eq. \ref{eq:permeable3rdInvIso} gives rise to vorticity $\mathbf{\omega}$ = curl $\mathbf{v}$ in the outflow velocity $\mathbf{v}$ interpreted as a current as illustrated in Figure \ref{chiralflow}.  Fluid flow carries momentum so the vorticity in the exit flow is observable. The spiral nature of the flow can be observed directly or via its consequences in reaction torque and force upon the porous material. 
\par
\begin{figure}[htb!]
\centering
\includegraphics[width=0.35\textwidth]{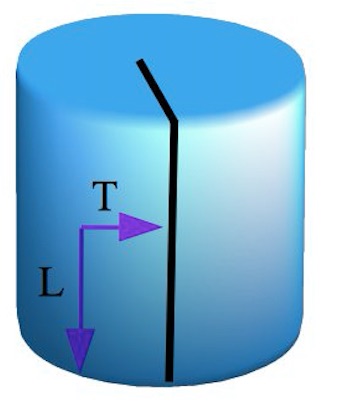} 
\caption{Fluid flow through a slotted chiral permeable cylinder with a vertical slot. Longitudinal flow (L) is unimpeded but transverse flow (T) due to chirality is interrupted by the slot, generating a pressure difference at the slot boundaries.}
\label{slotmethod1}
\end{figure}
The presence of transverse flow through a chiral cylinder can also be detected by preparing a cylinder with a vertical slot  (Figure \ref{slotmethod1}) with thin pressure sensors cemented to each wall of the slot. Any difference in pressure reveals transverse flow. 
\par
\begin{figure}[htb!]
\centering
\includegraphics[width=0.35\textwidth]{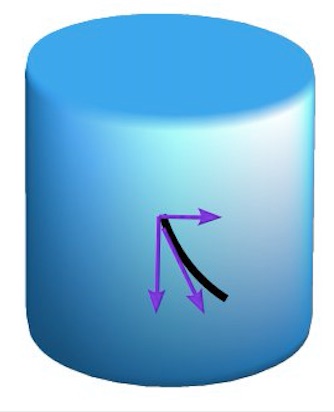} 
\caption{Fluid flow (arrows) through a slotted chiral permeable cylinder. A diagonal slot serves to detect the resultant direction (oblique arrow) of the flow. If the slot is aligned with the resultant flow direction then there will be no pressure difference on the boundaries of the slot.}
\label{slotmethod2}
\end{figure}

The characteristic length for flow can be inferred from the longitudinal flow velocity and the flow associated with exit vorticity. One may also infer the characteristic length by preparing a set of specimens with slots at different angles (Figure \ref{slotmethod2}). Thin wall pressure transducers on each slot wall reveal any differences in pressure due to transverse flow. The angle for which the pressure difference vanishes reveals  $\rho_{T} / \rho_{L}$ hence the characteristic length. 
\par
Heat flow, by contrast, carries negligible momentum associated with the phonons associated with motion of atoms or molecules. Upon exit from the chiral solid, the heat flow will have no macroscopically detectable characteristics. 

\section{Discussion}
As for comparison of fluid flow with heat flow, chiral, directionally isotropic gyroid lattices were observed experimentally to exhibit nonclassical thermal effects consistent with a third rank tensor property  \cite{Chiralheatflow25}. Specifically, a longitudinal temperature gradient was imposed on a cylinder of chiral gyroid lattice. Lattice specimens were heated at one end and were observed with an infrared camera.  A specimen was provided with a radial slot and compared with an intact specimen. Heat was observed to build up on one side of the slot, consistent with spiral flow of heat hence a third rank term. The lattices \cite{ReasaLakChiral20} had a surface wall thickness of 0.4 mm and a cell size of 6 mm. 
\par
As for comparison with second rank effects such as the possibility that the second rank conductivity tensor is asymmetric, consider that 
the heat flow in response to a radial temperature gradient in a disc has only a radial component from the classical conductivity tensor. The conductivity is expected to be symmetric \cite{Nye} but asymmetric tensor properties are known. For example the Hall effect can be analyzed via an asymmetric second rank tensor. Time reversal symmetry is broken by a magnetic field  \cite{Onsager2} so an asymmetric tensor is possible. The material may be homogeneous and isotropic. As for tests of possible asymmetry of the conductivity tensor, a slotted disc heated at the center can be used to test for such effects. Circulating flow due to an asymmetric conductivity tensor will be interrupted by the slot and there will be build up of heat along one edge. The method was tried with a single crystal of gypsum \cite{Soret1893} \cite{Soret1894} but no evidence of circulating flow was found.  Discs of gyroid lattice were studied more recently \cite{Chiralheatflow25} and no evidence of an asymmetric second rank tensor property was found. 
The present analysis and discussion have been confined to true tensors.  Pseudo-tensors have also been used in the analysis of chiral materials. 
\par
As for known asymmetric second rank tensors, wood has been observed to exhibit a fourth rank creep compliance tensor which is asymmetric with respect to exchange of pairs of indices  \cite{Neuhaus1983} \cite{Ozyha2013} \cite{Hering2012}. Wood is viscoelastic hence there is no conserved elastic energy density and time reversal invariance does not apply. Also, wood has orthotropic anisotropy.  Asymmetric modulus and compliance tensors in viscoelastic materials are anticipated theoretically \cite{Pipkin63}.
\par
In the present research, third rank permeability is predicted to reveal itself in vorticity of the fluid that exits the material. The swirling motion of the fluid is observable via optical methods. There will also be an observable reaction torque as well as a force upon the porous material. The ratio of these provides a way to infer the characteristic length.  For slow flow one may consider the slotted cylinder approach used for heat flow. A thin pressure sensor cemented to each surface of the slot will reveal differences in pressure due to the chirality. An oblique slot can be used to infer the ratio of transverse to longitudinal flow. If the pressure is equal on both sides of the oblique slot, then it is aligned with the direction of the flow.
\par
As for possible applications, effects of third rank permeability may be considered as a probe of the degree of chirality of heterogeneous materials. Materials of interest include the following. Surface lattices such as the gyroid as well as rib lattices are of interest. Ensemble averages of crystal shape in granules of sugar or quartz sand will have an effect on flow of air or liquid through a bed of granules. Beans and wheat in grain bins are expected to be chiral by virtue of their biological origin. Flow through beds of such grains can reveal average chirality. Similarly, flow of fluid through a container filled with wood screws can reveal their chirality. In a more speculative vein, flow through Martian sand could reveal chirality due to hypothetical biological processes.

\section{Conclusion}
Chirality allows a third rank tensor contribution to the permeability. This can give rise to vorticity of the outflow and to a reaction torque upon the permeable material. The higher rank freedom can be also expressed as part of an expansion of a nonlocal functional as has been done in elasticity, piezoelectricity and permeability.  As with Cosserat elasticity, the additional freedom can be expressed via one or more characteristic lengths. Methods for extracting a characteristic length from experiment are provided. 

\section{Acknowledgment}
I thank R. Bonazza for discussions on flow and R. L. Benedict for early collaboration on chirality in elasticity.\\\\


\begin{thebibliography}{30}  \setlength{\itemsep}{-1.0mm}
\bibitem{Nye} J. F. Nye, \emph{Physical Properties of Crystals}, Oxford, Clarendon, (1976). 
\bibitem{Lovett}  D. R. Lovett, \emph{Tensor Properties of Crystals}, Adam Hilger, Bristol and Philadelphia, (1989).
\bibitem{Onsager1}  L. Onsager, Reciprocal Relations in Irreversible Processes. I. Phys. Rev. 37, 405 (1931). 
\bibitem{Onsager2} L. Onsager, Reciprocal Relations in Irreversible Processes. II. Phys. Rev. 38, 2265 (1931).  
\bibitem{Sokolnikoff} Sokolnikoff, I. S., \emph{Theory of Elasticity}, Krieger; Malabar, FL, (1983).

\bibitem{Bursian1968}  Bursian, E., and N. Trunov, (1974), Nonlocal piezoelectric effect, Sov. Phys. Solid State - Fizika Tverdogo Tela 16(4), 1187-1190.
\bibitem{Bursian1974} Bursian, E., and O. I. Zaikovskii, (1968), Changes in the curvature of a ferroelectric film due to polarization, Sov. Phys. Solid State  - Fizika Tverdogo Tela 10, 1121-1124.
\bibitem{Trunov1975} N. Trunov, (1975), Polarization and susceptibility of a ferroelectric sample with a size comparable to the correlation radius, Sov. Phys. Solid State 17, 760-762.

\bibitem{EringenNonl72} Eringen, A. C., Linear theory of nonlocal elasticity and dispersion of plane waves, Int. J. Engng Sci, 10, 425-435, 1972.
\bibitem{WaismanNonl17} M. E. Mobasher, L. Berger- Vergiat, H. Waisman, Non-local formulation for transport and damage in porous media, Computer Methods in Applied Mechanics and Engineering 324,  654-688 (2017). 

\bibitem{Cosserat09}   E. Cosserat, and F. Cosserat, \textit{Theorie des Corps Deformables}, Hermann et Fils, Paris (1909).
\bibitem{Eringen68} A. C. Eringen,  Theory of micropolar elasticity. In Fracture Vol. \textbf{1}, 621-729 (edited by H. Liebowitz), Academic Press, New York (1968).
\bibitem{ParkLakesBone86} Park, H. C. and Lakes, R. S., Cosserat micromechanics of human bone:  strain redistribution by a hydration-sensitive constituent,  J. Biomechanics, 19 385-397 (1986).

\bibitem{NagawaDiamond80} M. Nagawa, K. Arakawa, M. Yamada, Diamond Crystals as Cosserat Continua, phys. stat. sol. (a)   57, 713-718 (1980)

\bibitem{RuLakesPhilM16}  Z. Rueger and R. S. Lakes, Experimental Cosserat elasticity in open cell polymer foam, Philosophical Magazine, 96 (2), 93-111, (2016).
\bibitem{RuegLak16}  Z. Rueger and  R. S. Lakes, Cosserat elasticity of negative Poisson's ratio foam: experiment, Smart Materials and Structures, \textbf{25}, 054004 (8pp) (2016).
\bibitem{RuegLakLatt18} Z. Rueger and  R. S. Lakes, Strong Cosserat elasticity in a transversely isotropic polymer lattice,  Phys. Rev. Lett., 120, 065501 Feb. (2018).

\bibitem{CossChiral82} R. S. Lakes, and R. L. Benedict, Noncentrosymmetry in micropolar elasticity. Int. J. of Engng. Sci. \textbf{20}, 1161 (1982).

\bibitem{HaLakChiral16} Ha, C. S., Plesha, M. E.,  Lakes, R. S.,   Chiral three-dimensional isotropic lattices with negative Poisson's ratio, Physica Status Solidi B, 253, (7), 1243-1251 (2016). 
\bibitem{ReasaLakChiral20} D. Reasa and R. S. Lakes, Chiral elasticity of the gyroid lattice,  Phys. Rev. Lett., 125, 205502,  (2020).
\bibitem{HaLakChiral20} Ha, C. S., Lakes, R. S., Plesha, M. E.,  Observation of squeeze twist coupling in a chiral three-dimensional isotropic lattice,  Physica Status Solidi B 257, (10), 1900140,  (2020).
\bibitem{GibAshby} Gibson, L. J. and Ashby, M. F., \textit{Cellular Solids}, Pergamon, Oxford; 2nd Ed., Cambridge (1997).

\bibitem{Soret1893}  C. Soret, De la conductibilite calorifique dans les cristaux. J. Phys. Theor. Appl., \textbf{2} (1), 241-259 (1893). 
\bibitem{Soret1894}   Ch. Saret, On the experimental investigation of the rotational coefficients of thermal conductivity, Philosophical Magazine Series 5, \textbf{37}, 226, 338-339, (1894).
\bibitem{Chiralheatflow25}  R. S. Lakes, Nonclassical heat flow in a passive chiral solid is third rank, not odd, Z. Angew. Math. Phys. (ZAMP)  76: 81 (2025). 

\bibitem{Neuhaus1983}  H. Neuhaus, Elastic behavior of spruce wood as a function of moisture content, Holz Roh-Werkst. \textbf{41}(1), 21-25 (1983).
\bibitem{Ozyha2013} T. Ozyhar, S. Hering, and P. Niemz, Viscoelastic characterization of wood: Time dependence of the orthotropic compliance in tension and compression, Journal of Rheology 57(2), 699-717 {2013}.
\bibitem{Hering2012} S. Hering, D. Keunecke, and P. Niemz, Moisture-dependent orthotropic elasticity of beech wood, Wood Sci. Technol. \textbf{46}, 927-938 (2012). 
\bibitem{Pipkin63}    A. C. Pipkin and G. Rogers, Asymmetric relaxation and compliance matrices in linear viscoelasticity, Z. Angew. Math. Phys. (ZAMP), \textbf{14}, 334-343 {1963}. 

\end{thebibliography}
\end{document}